\documentclass[aps,nobibnotes,prb,superscriptaddress,twocolumn,floatfix,showpacs,amsmath,amssymb, amsfonts,mathtools]{revtex4-1}
\usepackage{graphicx}
\usepackage[pdftex]{hyperref}
\usepackage{subcaption}
\usepackage{floatrow}
\usepackage{latexsym}
\usepackage{psfrag}
\usepackage[makeroom]{cancel}
\usepackage{dcolumn}
\usepackage{blindtext}
\usepackage{epstopdf}
\floatstyle{plaintop}
\restylefloat{table}

\usepackage{amsmath,latexsym,psfrag}
\usepackage{array}
\usepackage{dcolumn}
\usepackage{bm}
\usepackage{soul}  
\usepackage{gensymb}

\begin{document}
\title{ A decade of Density Functional Theory in Kenya}                    
\author{George S. Manyali}
	\affiliation{
		Computational and Theoretical Physics Group (CTheP), Department of Physical Sciences, Kaimosi Friends University, P.O Box 385-50309, Kaimosi, Kenya.}
\affiliation{Physics Department, Masinde Muliro University of Science and Technology,
          P.O Box 190-50100, Kakamega,Kenya}
	\date{\today}
\email[Corresponding Author:~]{gmanyali@kafu.ac.ke}

\begin{abstract}
 The African School Series on Electronic Structure Methods and Applications (ASESMA) has had a positive impact on growth of computational material science in Kenya, visibility of Kenyan universities and strong collaboration ties between Kenyan scientist and the rest of the world. Data retrieved from Scopus indicate that computational materials scientists in Kenya have published numerous articles, book chapters, and conference proceedings in peer-reviewed, high impact journals. The trend is likely to continue due to access to affordable computers, spread of electricity to remote areas, open-source softwares and cheaper internet. 
\setlength{\parskip}{1em}
\setlength{\parindent}{0pt}

\setlength{\parskip}{1em}
\setlength{\parindent}{0pt}

\end{abstract}
\maketitle

\section{Introduction}
The information extracted from Scopus database reveals that Kenyan scientists published over 40 thousands articles, book chapters, and conference proceedings in the last decade. This signifies an improvement in the quality of science in Kenya. Many of these articles published in high-ranking impact factor journals have attracted many citations. However, over 70\% of these projects are not funded by the government of Kenya. There are many possible reasons for this. Fundamentally, the funding agencies have their own priority areas to fund such as in applied sciences. By default, this eliminates projects undertaken in the field of basic sciences. Such funding structures suppresses creativity and innovation. Science should never be at the whims of funding bodies. To some extent, computational materials scientist through international collaborations have successfully demonstrated that there can never be the "Kenyan Science" or "African Science" or "Western Science". What matters is relevance of the science to the society bearing in mind that the world is a global village.

Scientists in Kenya should devise innovative ways to evade the challenges of limited funding to carry out a variety of science projects, to continue contributing to the knowledge database. The pooling of resources and collaborations with Africa and the Western continents is one such avenue that can bear bountiful fruits. Such networks are important in growing scientific disciplines based on theory and virtual laboratories. For instance, the computational materials scientists’ community is evolving rapidly due to availability of electricity throughout the country, computers, open-source codes, web-based data storage facilities, and availability of internet even in remote areas. Computational material science utilizes different simulation and modeling methods to expand our knowledge of different materials science topics. Through computation, we are able to get information of atomic or molecular interactions that are impossible to see through experimental methods. Density-functional theory (DFT) is one of the popular computational methods used in physics, chemistry and materials science used to investigate the electronic structure of many-body systems \citep{hohenberg1964inhomogeneous,kohn1965self}. There have been initiatives by international bodies to popularize computational materials science in Africa, and Kenya has not been left behind. The African School on Electronic Structure Methods and Applications (ASESMA) is an initiative that have contributed immensely on research activities surrounding DFT in Kenya and Africa at large 
\citep{Amolo201821,Amolo20182183,feder2011raising,hatt2013networking,chetty2020schooling,chetty2010material,chetty2011new}. Kenya hosted the 2012 ASESMA school which saw a number of western
scientists and others from Africa camping in the country for two weeks to discuss the basics of electronic structure and hands-on tutorials on simulations of materials. The seed sown at that time has led to growth of new crop of computational materials scientists equipped with both the theory and coding skills for development and application of DFT open-source codes\citep{Gómez-Ortiz2023,giannozzi2009quantum,giannozzi2017advanced,giannozzi2020quantum} 
. Their collaboration with western scientists have in principle enhanced the science in Kenyan institutions of higher learning and improved their visibility across the world. Today, Kenyan computational materials scientists have everything they need to do a world-class computational materials science research and compete favourable with  the rest of the world in generation on new ideas and expansion of existing knowledge about materials\citep{Yusuf2023,Ouma2012,Jomo2020,Odhiambo2018,MengWa201720359,MengWa201816765,Musembi2022,
Wafula2022a,Manyali2019,Makumi2023,Atambo2015,Wafula2023,Tindibale2023,Musembi2023,Fiorentin2020,Ongwen2023,Ongwen2021,Mbilo2022a,Mbilo2022b,
Muchiri2019489,Mengwa2016157,Kipkwarkwar2022,Mbilo2023252,Wafula2022b,
Ouma20184015,Shah2023,KingOri202030127,Korir2021,Omboga2020,Manyali2022,Philemon2023,
Korir2020,Ichibha2018168,Ouma2020,Vincent2021,Ongwen2022,Kingori2021,Ouma201725555,
Nyawere2014122,Ibraheem2017,Odhiambo201569,Odhiambo2015,Mbilo20232355,Kaner20235135,
Magnoungou2022,Namisi2023,Mbilo2022c,Agora2020,Mbae2022,Nyawere201425,Agora2022,
Motochi201210,Philemon2020,Muthui2017343,Chepkoech2018,Zipporah2018,Muthui2017343,
Nguimdo2016,Chepkoech2018,Zipporah2018}

\section*{Acknowledgements}
This work was inspired by lectures at ASESMA2023 school that took place in Kigali, Rwanda. The author is grateful for the financial support provided by Kenya Education Network (KENET) through Computational Modeling and Materials Science (CMMS) Research mini-grants 2019 and travel grant of June 2023. The computer resources at CTheP laboratory at Masinde Muliro University of Science and Technology and assistance provided by members of CTheP group are highly appreciated.


\bibliography{main.bib}

\begin{thebibliography}{70}%
\makeatletter
\providecommand \@ifxundefined [1]{%
 \@ifx{#1\undefined}
}%
\providecommand \@ifnum [1]{%
 \ifnum #1\expandafter \@firstoftwo
 \else \expandafter \@secondoftwo
 \fi
}%
\providecommand \@ifx [1]{%
 \ifx #1\expandafter \@firstoftwo
 \else \expandafter \@secondoftwo
 \fi
}%
\providecommand \natexlab [1]{#1}%
\providecommand \enquote  [1]{``#1''}%
\providecommand \bibnamefont  [1]{#1}%
\providecommand \bibfnamefont [1]{#1}%
\providecommand \citenamefont [1]{#1}%
\providecommand \href@noop [0]{\@secondoftwo}%
\providecommand \href [0]{\begingroup \@sanitize@url \@href}%
\providecommand \@href[1]{\@@startlink{#1}\@@href}%
\providecommand \@@href[1]{\endgroup#1\@@endlink}%
\providecommand \@sanitize@url [0]{\catcode `\\12\catcode `\$12\catcode
  `\&12\catcode `\#12\catcode `\^12\catcode `\_12\catcode `\%12\relax}%
\providecommand \@@startlink[1]{}%
\providecommand \@@endlink[0]{}%
\providecommand \url  [0]{\begingroup\@sanitize@url \@url }%
\providecommand \@url [1]{\endgroup\@href {#1}{\urlprefix }}%
\providecommand \urlprefix  [0]{URL }%
\providecommand \Eprint [0]{\href }%
\providecommand \doibase [0]{http://dx.doi.org/}%
\providecommand \selectlanguage [0]{\@gobble}%
\providecommand \bibinfo  [0]{\@secondoftwo}%
\providecommand \bibfield  [0]{\@secondoftwo}%
\providecommand \translation [1]{[#1]}%
\providecommand \BibitemOpen [0]{}%
\providecommand \bibitemStop [0]{}%
\providecommand \bibitemNoStop [0]{.\EOS\space}%
\providecommand \EOS [0]{\spacefactor3000\relax}%
\providecommand \BibitemShut  [1]{\csname bibitem#1\endcsname}%
\let\auto@bib@innerbib\@empty
\bibitem [{\citenamefont {Hohenberg}\ and\ \citenamefont
  {Kohn}(1964)}]{hohenberg1964inhomogeneous}%
  \BibitemOpen
  \bibfield  {author} {\bibinfo {author} {\bibfnamefont {P.}~\bibnamefont
  {Hohenberg}}\ and\ \bibinfo {author} {\bibfnamefont {W.}~\bibnamefont
  {Kohn}},\ }\href@noop {} {\bibfield  {journal} {\bibinfo  {journal} {Physical
  review}\ }\textbf {\bibinfo {volume} {136}},\ \bibinfo {pages} {B864}
  (\bibinfo {year} {1964})}\BibitemShut {NoStop}%
\bibitem [{\citenamefont {Kohn}\ and\ \citenamefont
  {Sham}(1965)}]{kohn1965self}%
  \BibitemOpen
  \bibfield  {author} {\bibinfo {author} {\bibfnamefont {W.}~\bibnamefont
  {Kohn}}\ and\ \bibinfo {author} {\bibfnamefont {L.~J.}\ \bibnamefont
  {Sham}},\ }\href@noop {} {\bibfield  {journal} {\bibinfo  {journal} {Physical
  review}\ }\textbf {\bibinfo {volume} {140}},\ \bibinfo {pages} {A1133}
  (\bibinfo {year} {1965})}\BibitemShut {NoStop}%
\bibitem [{\citenamefont {Amolo}(2018)}]{Amolo201821}%
  \BibitemOpen
  \bibfield  {author} {\bibinfo {author} {\bibfnamefont {G.~O.}\ \bibnamefont
  {Amolo}},\ }\href@noop {} {\bibfield  {journal} {\bibinfo  {journal}
  {Computing in Science and Engineering}\ }\textbf {\bibinfo {volume} {20}},\
  \bibinfo {pages} {21 – 24} (\bibinfo {year} {2018})}\BibitemShut {NoStop}%
\bibitem [{\citenamefont {Amolo}\ \emph {et~al.}(2018)\citenamefont {Amolo},
  \citenamefont {Chetty}, \citenamefont {Hassanali}, \citenamefont {Joubert},
  \citenamefont {Martin},\ and\ \citenamefont {Scandolo}}]{Amolo20182183}%
  \BibitemOpen
  \bibfield  {author} {\bibinfo {author} {\bibfnamefont {G.}~\bibnamefont
  {Amolo}}, \bibinfo {author} {\bibfnamefont {N.}~\bibnamefont {Chetty}},
  \bibinfo {author} {\bibfnamefont {A.}~\bibnamefont {Hassanali}}, \bibinfo
  {author} {\bibfnamefont {D.}~\bibnamefont {Joubert}}, \bibinfo {author}
  {\bibfnamefont {R.}~\bibnamefont {Martin}}, \ and\ \bibinfo {author}
  {\bibfnamefont {S.}~\bibnamefont {Scandolo}}\ }(\bibinfo {year} {2018})\ p.\
  \bibinfo {pages} {2183 – 2201}\BibitemShut {NoStop}%
\bibitem [{\citenamefont {Feder}(2011)}]{feder2011raising}%
  \BibitemOpen
  \bibfield  {author} {\bibinfo {author} {\bibfnamefont {T.}~\bibnamefont
  {Feder}},\ }\href@noop {} {\bibfield  {journal} {\bibinfo  {journal} {Physics
  Today}\ }\textbf {\bibinfo {volume} {64}},\ \bibinfo {pages} {28} (\bibinfo
  {year} {2011})}\BibitemShut {NoStop}%
\bibitem [{\citenamefont {Hatt}(2013)}]{hatt2013networking}%
  \BibitemOpen
  \bibfield  {author} {\bibinfo {author} {\bibfnamefont {A.}~\bibnamefont
  {Hatt}},\ }\href@noop {} {\bibfield  {journal} {\bibinfo  {journal} {MRS
  Bulletin}\ }\textbf {\bibinfo {volume} {38}},\ \bibinfo {pages} {12}
  (\bibinfo {year} {2013})}\BibitemShut {NoStop}%
\bibitem [{\citenamefont {Chetty}\ and\ \citenamefont
  {Martin}(2020)}]{chetty2020schooling}%
  \BibitemOpen
  \bibfield  {author} {\bibinfo {author} {\bibfnamefont {N.}~\bibnamefont
  {Chetty}}\ and\ \bibinfo {author} {\bibfnamefont {R.~M.}\ \bibnamefont
  {Martin}},\ }in\ \href@noop {} {\emph {\bibinfo {booktitle} {Journal of
  Physics: Conference Series}}},\ Vol.\ \bibinfo {volume} {1512}\ (\bibinfo
  {year} {2020})\ p.\ \bibinfo {pages} {012042}\BibitemShut {NoStop}%
\bibitem [{\citenamefont {Chetty}\ \emph {et~al.}(2010)\citenamefont {Chetty},
  \citenamefont {Martin},\ and\ \citenamefont {Scandolo}}]{chetty2010material}%
  \BibitemOpen
  \bibfield  {author} {\bibinfo {author} {\bibfnamefont {N.}~\bibnamefont
  {Chetty}}, \bibinfo {author} {\bibfnamefont {R.~M.}\ \bibnamefont {Martin}},
  \ and\ \bibinfo {author} {\bibfnamefont {S.}~\bibnamefont {Scandolo}},\
  }\href@noop {} {\bibfield  {journal} {\bibinfo  {journal} {Nature Physics}\
  }\textbf {\bibinfo {volume} {6}},\ \bibinfo {pages} {830} (\bibinfo {year}
  {2010})}\BibitemShut {NoStop}%
\bibitem [{\citenamefont {Chetty}(2011)}]{chetty2011new}%
  \BibitemOpen
  \bibfield  {author} {\bibinfo {author} {\bibfnamefont {N.}~\bibnamefont
  {Chetty}},\ }\href@noop {} {\bibfield  {journal} {\bibinfo  {journal}
  {Computer Physics Communications}\ }\textbf {\bibinfo {volume} {182}},\
  \bibinfo {pages} {2065} (\bibinfo {year} {2011})}\BibitemShut {NoStop}%
\bibitem [{\citenamefont {Gómez-Ortiz}\ \emph {et~al.}(2023)\citenamefont
  {Gómez-Ortiz}, \citenamefont {Carral-Sainz}, \citenamefont {Sifuna},
  \citenamefont {Monteseguro}, \citenamefont {Cuadrado}, \citenamefont
  {García-Fernández},\ and\ \citenamefont {Junquera}}]{Gómez-Ortiz2023}%
  \BibitemOpen
  \bibfield  {author} {\bibinfo {author} {\bibfnamefont {F.}~\bibnamefont
  {Gómez-Ortiz}}, \bibinfo {author} {\bibfnamefont {N.}~\bibnamefont
  {Carral-Sainz}}, \bibinfo {author} {\bibfnamefont {J.}~\bibnamefont
  {Sifuna}}, \bibinfo {author} {\bibfnamefont {V.}~\bibnamefont {Monteseguro}},
  \bibinfo {author} {\bibfnamefont {R.}~\bibnamefont {Cuadrado}}, \bibinfo
  {author} {\bibfnamefont {P.}~\bibnamefont {García-Fernández}}, \ and\
  \bibinfo {author} {\bibfnamefont {J.}~\bibnamefont {Junquera}},\ }\href@noop
  {} {\bibfield  {journal} {\bibinfo  {journal} {Computer Physics
  Communications}\ }\textbf {\bibinfo {volume} {286}} (\bibinfo {year}
  {2023})}\BibitemShut {NoStop}%
\bibitem [{\citenamefont {Giannozzi}\ \emph {et~al.}(2009)\citenamefont
  {Giannozzi}, \citenamefont {Baroni}, \citenamefont {Bonini}, \citenamefont
  {Calandra}, \citenamefont {Car}, \citenamefont {Cavazzoni}, \citenamefont
  {Ceresoli}, \citenamefont {Chiarotti}, \citenamefont {Cococcioni},
  \citenamefont {Dabo} \emph {et~al.}}]{giannozzi2009quantum}%
  \BibitemOpen
  \bibfield  {author} {\bibinfo {author} {\bibfnamefont {P.}~\bibnamefont
  {Giannozzi}}, \bibinfo {author} {\bibfnamefont {S.}~\bibnamefont {Baroni}},
  \bibinfo {author} {\bibfnamefont {N.}~\bibnamefont {Bonini}}, \bibinfo
  {author} {\bibfnamefont {M.}~\bibnamefont {Calandra}}, \bibinfo {author}
  {\bibfnamefont {R.}~\bibnamefont {Car}}, \bibinfo {author} {\bibfnamefont
  {C.}~\bibnamefont {Cavazzoni}}, \bibinfo {author} {\bibfnamefont
  {D.}~\bibnamefont {Ceresoli}}, \bibinfo {author} {\bibfnamefont {G.~L.}\
  \bibnamefont {Chiarotti}}, \bibinfo {author} {\bibfnamefont {M.}~\bibnamefont
  {Cococcioni}}, \bibinfo {author} {\bibfnamefont {I.}~\bibnamefont {Dabo}},
  \emph {et~al.},\ }\href@noop {} {\bibfield  {journal} {\bibinfo  {journal}
  {Journal of physics: Condensed matter}\ }\textbf {\bibinfo {volume} {21}},\
  \bibinfo {pages} {395502} (\bibinfo {year} {2009})}\BibitemShut {NoStop}%
\bibitem [{\citenamefont {Giannozzi}\ \emph {et~al.}(2017)\citenamefont
  {Giannozzi}, \citenamefont {Andreussi}, \citenamefont {Brumme}, \citenamefont
  {Bunau}, \citenamefont {Nardelli}, \citenamefont {Calandra}, \citenamefont
  {Car}, \citenamefont {Cavazzoni}, \citenamefont {Ceresoli}, \citenamefont
  {Cococcioni} \emph {et~al.}}]{giannozzi2017advanced}%
  \BibitemOpen
  \bibfield  {author} {\bibinfo {author} {\bibfnamefont {P.}~\bibnamefont
  {Giannozzi}}, \bibinfo {author} {\bibfnamefont {O.}~\bibnamefont
  {Andreussi}}, \bibinfo {author} {\bibfnamefont {T.}~\bibnamefont {Brumme}},
  \bibinfo {author} {\bibfnamefont {O.}~\bibnamefont {Bunau}}, \bibinfo
  {author} {\bibfnamefont {M.~B.}\ \bibnamefont {Nardelli}}, \bibinfo {author}
  {\bibfnamefont {M.}~\bibnamefont {Calandra}}, \bibinfo {author}
  {\bibfnamefont {R.}~\bibnamefont {Car}}, \bibinfo {author} {\bibfnamefont
  {C.}~\bibnamefont {Cavazzoni}}, \bibinfo {author} {\bibfnamefont
  {D.}~\bibnamefont {Ceresoli}}, \bibinfo {author} {\bibfnamefont
  {M.}~\bibnamefont {Cococcioni}},  \emph {et~al.},\ }\href@noop {} {\bibfield
  {journal} {\bibinfo  {journal} {Journal of physics: Condensed matter}\
  }\textbf {\bibinfo {volume} {29}},\ \bibinfo {pages} {465901} (\bibinfo
  {year} {2017})}\BibitemShut {NoStop}%
\bibitem [{\citenamefont {Giannozzi}\ \emph {et~al.}(2020)\citenamefont
  {Giannozzi}, \citenamefont {Baseggio}, \citenamefont {Bonf{\`a}},
  \citenamefont {Brunato}, \citenamefont {Car}, \citenamefont {Carnimeo},
  \citenamefont {Cavazzoni}, \citenamefont {De~Gironcoli}, \citenamefont
  {Delugas}, \citenamefont {Ferrari~Ruffino} \emph
  {et~al.}}]{giannozzi2020quantum}%
  \BibitemOpen
  \bibfield  {author} {\bibinfo {author} {\bibfnamefont {P.}~\bibnamefont
  {Giannozzi}}, \bibinfo {author} {\bibfnamefont {O.}~\bibnamefont {Baseggio}},
  \bibinfo {author} {\bibfnamefont {P.}~\bibnamefont {Bonf{\`a}}}, \bibinfo
  {author} {\bibfnamefont {D.}~\bibnamefont {Brunato}}, \bibinfo {author}
  {\bibfnamefont {R.}~\bibnamefont {Car}}, \bibinfo {author} {\bibfnamefont
  {I.}~\bibnamefont {Carnimeo}}, \bibinfo {author} {\bibfnamefont
  {C.}~\bibnamefont {Cavazzoni}}, \bibinfo {author} {\bibfnamefont
  {S.}~\bibnamefont {De~Gironcoli}}, \bibinfo {author} {\bibfnamefont
  {P.}~\bibnamefont {Delugas}}, \bibinfo {author} {\bibfnamefont
  {F.}~\bibnamefont {Ferrari~Ruffino}},  \emph {et~al.},\ }\href@noop {}
  {\bibfield  {journal} {\bibinfo  {journal} {The Journal of chemical physics}\
  }\textbf {\bibinfo {volume} {152}} (\bibinfo {year} {2020})}\BibitemShut
  {NoStop}%
\bibitem [{\citenamefont {Yusuf}\ \emph {et~al.}(2023)\citenamefont {Yusuf},
  \citenamefont {Saouma}, \citenamefont {Manyali}, \citenamefont {Wafula},\
  and\ \citenamefont {Huxley}}]{Yusuf2023}%
  \BibitemOpen
  \bibfield  {author} {\bibinfo {author} {\bibfnamefont {M.}~\bibnamefont
  {Yusuf}}, \bibinfo {author} {\bibfnamefont {F.~O.}\ \bibnamefont {Saouma}},
  \bibinfo {author} {\bibfnamefont {G.~S.}\ \bibnamefont {Manyali}}, \bibinfo
  {author} {\bibfnamefont {J.~W.}\ \bibnamefont {Wafula}}, \ and\ \bibinfo
  {author} {\bibfnamefont {O.}~\bibnamefont {Huxley}},\ }\href@noop {}
  {\bibfield  {journal} {\bibinfo  {journal} {Solid State Communications}\
  }\textbf {\bibinfo {volume} {370}} (\bibinfo {year} {2023})}\BibitemShut
  {NoStop}%
\bibitem [{\citenamefont {Ouma}\ \emph {et~al.}(2012)\citenamefont {Ouma},
  \citenamefont {Mapelu}, \citenamefont {Makau}, \citenamefont {Amolo},\ and\
  \citenamefont {Maezono}}]{Ouma2012}%
  \BibitemOpen
  \bibfield  {author} {\bibinfo {author} {\bibfnamefont {C.}~\bibnamefont
  {Ouma}}, \bibinfo {author} {\bibfnamefont {M.}~\bibnamefont {Mapelu}},
  \bibinfo {author} {\bibfnamefont {N.}~\bibnamefont {Makau}}, \bibinfo
  {author} {\bibfnamefont {G.}~\bibnamefont {Amolo}}, \ and\ \bibinfo {author}
  {\bibfnamefont {R.}~\bibnamefont {Maezono}},\ }\href@noop {} {\bibfield
  {journal} {\bibinfo  {journal} {Physical Review B - Condensed Matter and
  Materials Physics}\ }\textbf {\bibinfo {volume} {86}} (\bibinfo {year}
  {2012})}\BibitemShut {NoStop}%
\bibitem [{\citenamefont {Jomo}\ \emph {et~al.}(2020)\citenamefont {Jomo},
  \citenamefont {Otieno},\ and\ \citenamefont {Nyawere}}]{Jomo2020}%
  \BibitemOpen
  \bibfield  {author} {\bibinfo {author} {\bibfnamefont {P.}~\bibnamefont
  {Jomo}}, \bibinfo {author} {\bibfnamefont {C.}~\bibnamefont {Otieno}}, \ and\
  \bibinfo {author} {\bibfnamefont {P.}~\bibnamefont {Nyawere}},\ }\href@noop
  {} {\bibfield  {journal} {\bibinfo  {journal} {Advances in Condensed Matter
  Physics}\ }\textbf {\bibinfo {volume} {2020}} (\bibinfo {year}
  {2020})}\BibitemShut {NoStop}%
\bibitem [{\citenamefont {Odhiambo}\ \emph {et~al.}(2018)\citenamefont
  {Odhiambo}, \citenamefont {Amolo}, \citenamefont {Makau}, \citenamefont
  {Kiptiemoi},\ and\ \citenamefont {Othieno}}]{Odhiambo2018}%
  \BibitemOpen
  \bibfield  {author} {\bibinfo {author} {\bibfnamefont {H.}~\bibnamefont
  {Odhiambo}}, \bibinfo {author} {\bibfnamefont {G.}~\bibnamefont {Amolo}},
  \bibinfo {author} {\bibfnamefont {N.}~\bibnamefont {Makau}}, \bibinfo
  {author} {\bibfnamefont {K.}~\bibnamefont {Kiptiemoi}}, \ and\ \bibinfo
  {author} {\bibfnamefont {H.}~\bibnamefont {Othieno}},\ }\href@noop {}
  {\bibfield  {journal} {\bibinfo  {journal} {International Journal of
  Nanoscience}\ }\textbf {\bibinfo {volume} {17}} (\bibinfo {year}
  {2018})}\BibitemShut {NoStop}%
\bibitem [{\citenamefont {Meng'Wa}\ \emph {et~al.}(2017)\citenamefont
  {Meng'Wa}, \citenamefont {Makau}, \citenamefont {Amolo}, \citenamefont
  {Scandolo},\ and\ \citenamefont {Seriani}}]{MengWa201720359}%
  \BibitemOpen
  \bibfield  {author} {\bibinfo {author} {\bibfnamefont {V.}~\bibnamefont
  {Meng'Wa}}, \bibinfo {author} {\bibfnamefont {N.}~\bibnamefont {Makau}},
  \bibinfo {author} {\bibfnamefont {G.}~\bibnamefont {Amolo}}, \bibinfo
  {author} {\bibfnamefont {S.}~\bibnamefont {Scandolo}}, \ and\ \bibinfo
  {author} {\bibfnamefont {N.}~\bibnamefont {Seriani}},\ }\href@noop {}
  {\bibfield  {journal} {\bibinfo  {journal} {Journal of Physical Chemistry C}\
  }\textbf {\bibinfo {volume} {121}},\ \bibinfo {pages} {20359 – 20365}
  (\bibinfo {year} {2017})}\BibitemShut {NoStop}%
\bibitem [{\citenamefont {Meng'Wa}\ \emph {et~al.}(2018)\citenamefont
  {Meng'Wa}, \citenamefont {Makau}, \citenamefont {Amolo}, \citenamefont
  {Scandolo},\ and\ \citenamefont {Seriani}}]{MengWa201816765}%
  \BibitemOpen
  \bibfield  {author} {\bibinfo {author} {\bibfnamefont {V.}~\bibnamefont
  {Meng'Wa}}, \bibinfo {author} {\bibfnamefont {N.}~\bibnamefont {Makau}},
  \bibinfo {author} {\bibfnamefont {G.}~\bibnamefont {Amolo}}, \bibinfo
  {author} {\bibfnamefont {S.}~\bibnamefont {Scandolo}}, \ and\ \bibinfo
  {author} {\bibfnamefont {N.}~\bibnamefont {Seriani}},\ }\href@noop {}
  {\bibfield  {journal} {\bibinfo  {journal} {Journal of Physical Chemistry C}\
  }\textbf {\bibinfo {volume} {122}},\ \bibinfo {pages} {16765 – 16771}
  (\bibinfo {year} {2018})}\BibitemShut {NoStop}%
\bibitem [{\citenamefont {Musembi}\ and\ \citenamefont
  {Mbilo}(2022)}]{Musembi2022}%
  \BibitemOpen
  \bibfield  {author} {\bibinfo {author} {\bibfnamefont {R.}~\bibnamefont
  {Musembi}}\ and\ \bibinfo {author} {\bibfnamefont {M.}~\bibnamefont
  {Mbilo}},\ }\href@noop {} {\bibfield  {journal} {\bibinfo  {journal}
  {Materialia}\ }\textbf {\bibinfo {volume} {26}} (\bibinfo {year}
  {2022})}\BibitemShut {NoStop}%
\bibitem [{\citenamefont {Wafula}\ \emph
  {et~al.}(2022{\natexlab{a}})\citenamefont {Wafula}, \citenamefont {Makokha},\
  and\ \citenamefont {Manyali}}]{Wafula2022a}%
  \BibitemOpen
  \bibfield  {author} {\bibinfo {author} {\bibfnamefont {J.~W.}\ \bibnamefont
  {Wafula}}, \bibinfo {author} {\bibfnamefont {J.~W.}\ \bibnamefont {Makokha}},
  \ and\ \bibinfo {author} {\bibfnamefont {G.~S.}\ \bibnamefont {Manyali}},\
  }\href@noop {} {\bibfield  {journal} {\bibinfo  {journal} {Results in
  Physics}\ }\textbf {\bibinfo {volume} {43}} (\bibinfo {year}
  {2022}{\natexlab{a}})}\BibitemShut {NoStop}%
\bibitem [{\citenamefont {Manyali}\ and\ \citenamefont
  {Sifuna}(2019)}]{Manyali2019}%
  \BibitemOpen
  \bibfield  {author} {\bibinfo {author} {\bibfnamefont {G.~S.}\ \bibnamefont
  {Manyali}}\ and\ \bibinfo {author} {\bibfnamefont {J.}~\bibnamefont
  {Sifuna}},\ }\href@noop {} {\bibfield  {journal} {\bibinfo  {journal} {AIP
  Advances}\ }\textbf {\bibinfo {volume} {9}} (\bibinfo {year}
  {2019})}\BibitemShut {NoStop}%
\bibitem [{\citenamefont {Makumi}\ \emph {et~al.}(2023)\citenamefont {Makumi},
  \citenamefont {Bem}, \citenamefont {Musila}, \citenamefont {Foss},\ and\
  \citenamefont {Aksamija}}]{Makumi2023}%
  \BibitemOpen
  \bibfield  {author} {\bibinfo {author} {\bibfnamefont {S.~W.}\ \bibnamefont
  {Makumi}}, \bibinfo {author} {\bibfnamefont {D.}~\bibnamefont {Bem}},
  \bibinfo {author} {\bibfnamefont {N.}~\bibnamefont {Musila}}, \bibinfo
  {author} {\bibfnamefont {C.}~\bibnamefont {Foss}}, \ and\ \bibinfo {author}
  {\bibfnamefont {Z.}~\bibnamefont {Aksamija}},\ }\href@noop {} {\bibfield
  {journal} {\bibinfo  {journal} {Journal of Physics Condensed Matter}\
  }\textbf {\bibinfo {volume} {35}} (\bibinfo {year} {2023})}\BibitemShut
  {NoStop}%
\bibitem [{\citenamefont {Atambo}\ \emph {et~al.}(2015)\citenamefont {Atambo},
  \citenamefont {Makau}, \citenamefont {Amolo},\ and\ \citenamefont
  {Maezono}}]{Atambo2015}%
  \BibitemOpen
  \bibfield  {author} {\bibinfo {author} {\bibfnamefont {M.~O.}\ \bibnamefont
  {Atambo}}, \bibinfo {author} {\bibfnamefont {N.}~\bibnamefont {Makau}},
  \bibinfo {author} {\bibfnamefont {G.}~\bibnamefont {Amolo}}, \ and\ \bibinfo
  {author} {\bibfnamefont {R.}~\bibnamefont {Maezono}},\ }\href@noop {}
  {\bibfield  {journal} {\bibinfo  {journal} {Materials Research Express}\
  }\textbf {\bibinfo {volume} {2}} (\bibinfo {year} {2015})}\BibitemShut
  {NoStop}%
\bibitem [{\citenamefont {Wafula}(2023)}]{Wafula2023}%
  \BibitemOpen
  \bibfield  {author} {\bibinfo {author} {\bibfnamefont {J.~W.}\ \bibnamefont
  {Wafula}},\ }\href@noop {} {\bibfield  {journal} {\bibinfo  {journal}
  {Results in Materials}\ }\textbf {\bibinfo {volume} {19}} (\bibinfo {year}
  {2023})}\BibitemShut {NoStop}%
\bibitem [{\citenamefont {Tindibale}\ \emph {et~al.}(2023)\citenamefont
  {Tindibale}, \citenamefont {Mulwa},\ and\ \citenamefont
  {Adetunji}}]{Tindibale2023}%
  \BibitemOpen
  \bibfield  {author} {\bibinfo {author} {\bibfnamefont {E.}~\bibnamefont
  {Tindibale}}, \bibinfo {author} {\bibfnamefont {W.~M.}\ \bibnamefont
  {Mulwa}}, \ and\ \bibinfo {author} {\bibfnamefont {B.~I.}\ \bibnamefont
  {Adetunji}},\ }\href@noop {} {\bibfield  {journal} {\bibinfo  {journal}
  {Physica B: Condensed Matter}\ }\textbf {\bibinfo {volume} {665}} (\bibinfo
  {year} {2023})}\BibitemShut {NoStop}%
\bibitem [{\citenamefont {Musembi}\ and\ \citenamefont
  {Mbilo}(2023)}]{Musembi2023}%
  \BibitemOpen
  \bibfield  {author} {\bibinfo {author} {\bibfnamefont {R.}~\bibnamefont
  {Musembi}}\ and\ \bibinfo {author} {\bibfnamefont {M.}~\bibnamefont
  {Mbilo}},\ }\href@noop {} {\bibfield  {journal} {\bibinfo  {journal} {MRS
  Advances}\ } (\bibinfo {year} {2023})}\BibitemShut {NoStop}%
\bibitem [{\citenamefont {Re~Fiorentin}\ \emph {et~al.}(2020)\citenamefont
  {Re~Fiorentin}, \citenamefont {Kiprono},\ and\ \citenamefont
  {Risplendi}}]{Fiorentin2020}%
  \BibitemOpen
  \bibfield  {author} {\bibinfo {author} {\bibfnamefont {M.}~\bibnamefont
  {Re~Fiorentin}}, \bibinfo {author} {\bibfnamefont {K.~K.}\ \bibnamefont
  {Kiprono}}, \ and\ \bibinfo {author} {\bibfnamefont {F.}~\bibnamefont
  {Risplendi}},\ }\href@noop {} {\bibfield  {journal} {\bibinfo  {journal}
  {Nanomaterials and Nanotechnology}\ }\textbf {\bibinfo {volume} {10}}
  (\bibinfo {year} {2020})}\BibitemShut {NoStop}%
\bibitem [{\citenamefont {Ongwen}\ \emph {et~al.}(2023)\citenamefont {Ongwen},
  \citenamefont {Ogam}, \citenamefont {Fellah}, \citenamefont {Mageto},
  \citenamefont {Othieno},\ and\ \citenamefont {Otunga}}]{Ongwen2023}%
  \BibitemOpen
  \bibfield  {author} {\bibinfo {author} {\bibfnamefont {N.}~\bibnamefont
  {Ongwen}}, \bibinfo {author} {\bibfnamefont {E.}~\bibnamefont {Ogam}},
  \bibinfo {author} {\bibfnamefont {Z.}~\bibnamefont {Fellah}}, \bibinfo
  {author} {\bibfnamefont {M.}~\bibnamefont {Mageto}}, \bibinfo {author}
  {\bibfnamefont {H.}~\bibnamefont {Othieno}}, \ and\ \bibinfo {author}
  {\bibfnamefont {H.}~\bibnamefont {Otunga}},\ }\href@noop {} {\bibfield
  {journal} {\bibinfo  {journal} {Physica B: Condensed Matter}\ }\textbf
  {\bibinfo {volume} {651}} (\bibinfo {year} {2023})}\BibitemShut {NoStop}%
\bibitem [{\citenamefont {Ongwen}\ \emph {et~al.}(2021)\citenamefont {Ongwen},
  \citenamefont {Ogam},\ and\ \citenamefont {Otunga}}]{Ongwen2021}%
  \BibitemOpen
  \bibfield  {author} {\bibinfo {author} {\bibfnamefont {N.~O.}\ \bibnamefont
  {Ongwen}}, \bibinfo {author} {\bibfnamefont {E.}~\bibnamefont {Ogam}}, \ and\
  \bibinfo {author} {\bibfnamefont {H.~O.}\ \bibnamefont {Otunga}},\
  }\href@noop {} {\bibfield  {journal} {\bibinfo  {journal} {Materials Today
  Communications}\ }\textbf {\bibinfo {volume} {26}} (\bibinfo {year}
  {2021})}\BibitemShut {NoStop}%
\bibitem [{\citenamefont {Mbilo}\ \emph {et~al.}(2022)\citenamefont {Mbilo},
  \citenamefont {Manyali},\ and\ \citenamefont {Musembi}}]{Mbilo2022a}%
  \BibitemOpen
  \bibfield  {author} {\bibinfo {author} {\bibfnamefont {M.}~\bibnamefont
  {Mbilo}}, \bibinfo {author} {\bibfnamefont {G.~S.}\ \bibnamefont {Manyali}},
  \ and\ \bibinfo {author} {\bibfnamefont {R.~J.}\ \bibnamefont {Musembi}},\
  }\href@noop {} {\bibfield  {journal} {\bibinfo  {journal} {Computational
  Condensed Matter}\ }\textbf {\bibinfo {volume} {32}} (\bibinfo {year}
  {2022})}\BibitemShut {NoStop}%
\bibitem [{\citenamefont {Mbilo}\ and\ \citenamefont
  {Musembi}(2022{\natexlab{a}})}]{Mbilo2022b}%
  \BibitemOpen
  \bibfield  {author} {\bibinfo {author} {\bibfnamefont {M.}~\bibnamefont
  {Mbilo}}\ and\ \bibinfo {author} {\bibfnamefont {R.}~\bibnamefont
  {Musembi}},\ }\href@noop {} {\bibfield  {journal} {\bibinfo  {journal}
  {Advances in Materials Science and Engineering}\ }\textbf {\bibinfo {volume}
  {2022}} (\bibinfo {year} {2022}{\natexlab{a}})}\BibitemShut {NoStop}%
\bibitem [{\citenamefont {Muchiri}\ \emph {et~al.}(2019)\citenamefont
  {Muchiri}, \citenamefont {Mwalukuku}, \citenamefont {Korir}, \citenamefont
  {Amolo},\ and\ \citenamefont {Makau}}]{Muchiri2019489}%
  \BibitemOpen
  \bibfield  {author} {\bibinfo {author} {\bibfnamefont {P.}~\bibnamefont
  {Muchiri}}, \bibinfo {author} {\bibfnamefont {V.}~\bibnamefont {Mwalukuku}},
  \bibinfo {author} {\bibfnamefont {K.}~\bibnamefont {Korir}}, \bibinfo
  {author} {\bibfnamefont {G.}~\bibnamefont {Amolo}}, \ and\ \bibinfo {author}
  {\bibfnamefont {N.}~\bibnamefont {Makau}},\ }\href@noop {} {\bibfield
  {journal} {\bibinfo  {journal} {Materials Chemistry and Physics}\ }\textbf
  {\bibinfo {volume} {229}},\ \bibinfo {pages} {489 – 494} (\bibinfo {year}
  {2019})}\BibitemShut {NoStop}%
\bibitem [{\citenamefont {Meng'wa}\ \emph {et~al.}(2016)\citenamefont
  {Meng'wa}, \citenamefont {Amolo}, \citenamefont {Makau}, \citenamefont
  {Lutta}, \citenamefont {Okoth}, \citenamefont {Mwabora}, \citenamefont
  {Musembi}, \citenamefont {Maghanga},\ and\ \citenamefont
  {Gateru}}]{Mengwa2016157}%
  \BibitemOpen
  \bibfield  {author} {\bibinfo {author} {\bibfnamefont {V.}~\bibnamefont
  {Meng'wa}}, \bibinfo {author} {\bibfnamefont {G.}~\bibnamefont {Amolo}},
  \bibinfo {author} {\bibfnamefont {N.}~\bibnamefont {Makau}}, \bibinfo
  {author} {\bibfnamefont {S.}~\bibnamefont {Lutta}}, \bibinfo {author}
  {\bibfnamefont {M.}~\bibnamefont {Okoth}}, \bibinfo {author} {\bibfnamefont
  {J.}~\bibnamefont {Mwabora}}, \bibinfo {author} {\bibfnamefont
  {R.}~\bibnamefont {Musembi}}, \bibinfo {author} {\bibfnamefont
  {C.}~\bibnamefont {Maghanga}}, \ and\ \bibinfo {author} {\bibfnamefont
  {R.}~\bibnamefont {Gateru}},\ }\href@noop {} {\bibfield  {journal} {\bibinfo
  {journal} {African Review of Physics}\ }\textbf {\bibinfo {volume} {11}},\
  \bibinfo {pages} {157 – 165} (\bibinfo {year} {2016})}\BibitemShut
  {NoStop}%
\bibitem [{\citenamefont {Kipkwarkwar}\ \emph {et~al.}(2022)\citenamefont
  {Kipkwarkwar}, \citenamefont {Nyawere},\ and\ \citenamefont
  {Maghanga}}]{Kipkwarkwar2022}%
  \BibitemOpen
  \bibfield  {author} {\bibinfo {author} {\bibfnamefont {T.~J.}\ \bibnamefont
  {Kipkwarkwar}}, \bibinfo {author} {\bibfnamefont {P.}~\bibnamefont
  {Nyawere}}, \ and\ \bibinfo {author} {\bibfnamefont {C.}~\bibnamefont
  {Maghanga}},\ }\href@noop {} {\bibfield  {journal} {\bibinfo  {journal}
  {Advances in Condensed Matter Physics}\ }\textbf {\bibinfo {volume} {2022}}
  (\bibinfo {year} {2022})}\BibitemShut {NoStop}%
\bibitem [{\citenamefont {Mbilo}\ and\ \citenamefont
  {Musembi}(2023)}]{Mbilo2023252}%
  \BibitemOpen
  \bibfield  {author} {\bibinfo {author} {\bibfnamefont {M.}~\bibnamefont
  {Mbilo}}\ and\ \bibinfo {author} {\bibfnamefont {R.}~\bibnamefont
  {Musembi}},\ }\href@noop {} {\bibfield  {journal} {\bibinfo  {journal}
  {Journal of the Korean Ceramic Society}\ }\textbf {\bibinfo {volume} {60}},\
  \bibinfo {pages} {252 – 260} (\bibinfo {year} {2023})}\BibitemShut
  {NoStop}%
\bibitem [{\citenamefont {Wafula}\ \emph
  {et~al.}(2022{\natexlab{b}})\citenamefont {Wafula}, \citenamefont {Manyali},\
  and\ \citenamefont {Makokha}}]{Wafula2022b}%
  \BibitemOpen
  \bibfield  {author} {\bibinfo {author} {\bibfnamefont {J.~W.}\ \bibnamefont
  {Wafula}}, \bibinfo {author} {\bibfnamefont {G.~S.}\ \bibnamefont {Manyali}},
  \ and\ \bibinfo {author} {\bibfnamefont {J.~W.}\ \bibnamefont {Makokha}},\
  }\href@noop {} {\bibfield  {journal} {\bibinfo  {journal} {Oxford Open
  Materials Science}\ }\textbf {\bibinfo {volume} {2}} (\bibinfo {year}
  {2022}{\natexlab{b}})}\BibitemShut {NoStop}%
\bibitem [{\citenamefont {Ouma}\ \emph {et~al.}(2018)\citenamefont {Ouma},
  \citenamefont {Obodo}, \citenamefont {Braun},\ and\ \citenamefont
  {Amolo}}]{Ouma20184015}%
  \BibitemOpen
  \bibfield  {author} {\bibinfo {author} {\bibfnamefont {C.~N.~M.}\
  \bibnamefont {Ouma}}, \bibinfo {author} {\bibfnamefont {K.~O.}\ \bibnamefont
  {Obodo}}, \bibinfo {author} {\bibfnamefont {M.}~\bibnamefont {Braun}}, \ and\
  \bibinfo {author} {\bibfnamefont {G.~O.}\ \bibnamefont {Amolo}},\ }\href@noop
  {} {\bibfield  {journal} {\bibinfo  {journal} {Journal of Materials Chemistry
  C}\ }\textbf {\bibinfo {volume} {6}},\ \bibinfo {pages} {4015 – 4022}
  (\bibinfo {year} {2018})}\BibitemShut {NoStop}%
\bibitem [{\citenamefont {Zuhair Abbas~Shah}\ \emph {et~al.}(2023)\citenamefont
  {Zuhair Abbas~Shah}, \citenamefont {Niaz}, \citenamefont {Nasir},\ and\
  \citenamefont {Sifuna}}]{Shah2023}%
  \BibitemOpen
  \bibfield  {author} {\bibinfo {author} {\bibfnamefont {S.}~\bibnamefont
  {Zuhair Abbas~Shah}}, \bibinfo {author} {\bibfnamefont {S.}~\bibnamefont
  {Niaz}}, \bibinfo {author} {\bibfnamefont {T.}~\bibnamefont {Nasir}}, \ and\
  \bibinfo {author} {\bibfnamefont {J.}~\bibnamefont {Sifuna}},\ }\href@noop {}
  {\bibfield  {journal} {\bibinfo  {journal} {Results in Chemistry}\ }\textbf
  {\bibinfo {volume} {5}} (\bibinfo {year} {2023})}\BibitemShut {NoStop}%
\bibitem [{\citenamefont {King'Ori}\ \emph {et~al.}(2020)\citenamefont
  {King'Ori}, \citenamefont {Ouma}, \citenamefont {Mishra}, \citenamefont
  {Amolo},\ and\ \citenamefont {Makau}}]{KingOri202030127}%
  \BibitemOpen
  \bibfield  {author} {\bibinfo {author} {\bibfnamefont {G.~W.}\ \bibnamefont
  {King'Ori}}, \bibinfo {author} {\bibfnamefont {C.~N.~M.}\ \bibnamefont
  {Ouma}}, \bibinfo {author} {\bibfnamefont {A.~K.}\ \bibnamefont {Mishra}},
  \bibinfo {author} {\bibfnamefont {G.~O.}\ \bibnamefont {Amolo}}, \ and\
  \bibinfo {author} {\bibfnamefont {N.~W.}\ \bibnamefont {Makau}},\ }\href@noop
  {} {\bibfield  {journal} {\bibinfo  {journal} {RSC Advances}\ }\textbf
  {\bibinfo {volume} {10}},\ \bibinfo {pages} {30127 – 30138} (\bibinfo
  {year} {2020})}\BibitemShut {NoStop}%
\bibitem [{\citenamefont {Korir}\ \emph {et~al.}(2021)\citenamefont {Korir},
  \citenamefont {Benecha}, \citenamefont {Nyamwala},\ and\ \citenamefont
  {Lombardi}}]{Korir2021}%
  \BibitemOpen
  \bibfield  {author} {\bibinfo {author} {\bibfnamefont {K.}~\bibnamefont
  {Korir}}, \bibinfo {author} {\bibfnamefont {E.}~\bibnamefont {Benecha}},
  \bibinfo {author} {\bibfnamefont {F.}~\bibnamefont {Nyamwala}}, \ and\
  \bibinfo {author} {\bibfnamefont {E.}~\bibnamefont {Lombardi}},\ }\href@noop
  {} {\bibfield  {journal} {\bibinfo  {journal} {Materials Today
  Communications}\ }\textbf {\bibinfo {volume} {26}} (\bibinfo {year}
  {2021})}\BibitemShut {NoStop}%
\bibitem [{\citenamefont {Omboga}\ and\ \citenamefont
  {Otieno}(2020)}]{Omboga2020}%
  \BibitemOpen
  \bibfield  {author} {\bibinfo {author} {\bibfnamefont {N.}~\bibnamefont
  {Omboga}}\ and\ \bibinfo {author} {\bibfnamefont {C.}~\bibnamefont
  {Otieno}},\ }\href@noop {} {\bibfield  {journal} {\bibinfo  {journal}
  {Journal of Physics Communications}\ }\textbf {\bibinfo {volume} {4}}
  (\bibinfo {year} {2020})}\BibitemShut {NoStop}%
\bibitem [{\citenamefont {Manyali}(2022)}]{Manyali2022}%
  \BibitemOpen
  \bibfield  {author} {\bibinfo {author} {\bibfnamefont {G.~S.}\ \bibnamefont
  {Manyali}},\ }\href@noop {} {\bibfield  {journal} {\bibinfo  {journal}
  {Materials Today Communications}\ }\textbf {\bibinfo {volume} {30}} (\bibinfo
  {year} {2022})}\BibitemShut {NoStop}%
\bibitem [{\citenamefont {Philemon}\ \emph {et~al.}(2023)\citenamefont
  {Philemon}, \citenamefont {Korir}, \citenamefont {Musembi},\ and\
  \citenamefont {Nyongesa}}]{Philemon2023}%
  \BibitemOpen
  \bibfield  {author} {\bibinfo {author} {\bibfnamefont {K.~T.}\ \bibnamefont
  {Philemon}}, \bibinfo {author} {\bibfnamefont {K.~K.}\ \bibnamefont {Korir}},
  \bibinfo {author} {\bibfnamefont {R.~J.}\ \bibnamefont {Musembi}}, \ and\
  \bibinfo {author} {\bibfnamefont {F.~W.}\ \bibnamefont {Nyongesa}},\
  }\href@noop {} {\bibfield  {journal} {\bibinfo  {journal} {Materialia}\
  }\textbf {\bibinfo {volume} {29}} (\bibinfo {year} {2023})}\BibitemShut
  {NoStop}%
\bibitem [{\citenamefont {Korir}\ and\ \citenamefont
  {Philemon}(2020)}]{Korir2020}%
  \BibitemOpen
  \bibfield  {author} {\bibinfo {author} {\bibfnamefont {K.}~\bibnamefont
  {Korir}}\ and\ \bibinfo {author} {\bibfnamefont {K.~T.}\ \bibnamefont
  {Philemon}},\ }\href@noop {} {\bibfield  {journal} {\bibinfo  {journal}
  {Materialia}\ }\textbf {\bibinfo {volume} {11}} (\bibinfo {year}
  {2020})}\BibitemShut {NoStop}%
\bibitem [{\citenamefont {Ichibha}\ \emph {et~al.}(2018)\citenamefont
  {Ichibha}, \citenamefont {Hongo}, \citenamefont {Motochi}, \citenamefont
  {Makau}, \citenamefont {Amolo},\ and\ \citenamefont
  {Maezono}}]{Ichibha2018168}%
  \BibitemOpen
  \bibfield  {author} {\bibinfo {author} {\bibfnamefont {T.}~\bibnamefont
  {Ichibha}}, \bibinfo {author} {\bibfnamefont {K.}~\bibnamefont {Hongo}},
  \bibinfo {author} {\bibfnamefont {I.}~\bibnamefont {Motochi}}, \bibinfo
  {author} {\bibfnamefont {N.}~\bibnamefont {Makau}}, \bibinfo {author}
  {\bibfnamefont {G.}~\bibnamefont {Amolo}}, \ and\ \bibinfo {author}
  {\bibfnamefont {R.}~\bibnamefont {Maezono}},\ }\href@noop {} {\bibfield
  {journal} {\bibinfo  {journal} {Diamond and Related Materials}\ }\textbf
  {\bibinfo {volume} {81}},\ \bibinfo {pages} {168 – 175} (\bibinfo {year}
  {2018})}\BibitemShut {NoStop}%
\bibitem [{\citenamefont {Ouma}\ \emph {et~al.}(2020)\citenamefont {Ouma},
  \citenamefont {Obodo}, \citenamefont {Parlak},\ and\ \citenamefont
  {Amolo}}]{Ouma2020}%
  \BibitemOpen
  \bibfield  {author} {\bibinfo {author} {\bibfnamefont {C.~N.}\ \bibnamefont
  {Ouma}}, \bibinfo {author} {\bibfnamefont {K.~O.}\ \bibnamefont {Obodo}},
  \bibinfo {author} {\bibfnamefont {C.}~\bibnamefont {Parlak}}, \ and\ \bibinfo
  {author} {\bibfnamefont {G.~O.}\ \bibnamefont {Amolo}},\ }\href@noop {}
  {\bibfield  {journal} {\bibinfo  {journal} {Physica E: Low-Dimensional
  Systems and Nanostructures}\ }\textbf {\bibinfo {volume} {123}} (\bibinfo
  {year} {2020})}\BibitemShut {NoStop}%
\bibitem [{\citenamefont {Vincent}\ \emph {et~al.}(2021)\citenamefont
  {Vincent}, \citenamefont {Mulwa},\ and\ \citenamefont {Kirui}}]{Vincent2021}%
  \BibitemOpen
  \bibfield  {author} {\bibinfo {author} {\bibfnamefont {O.}~\bibnamefont
  {Vincent}}, \bibinfo {author} {\bibfnamefont {W.~M.}\ \bibnamefont {Mulwa}},
  \ and\ \bibinfo {author} {\bibfnamefont {M.}~\bibnamefont {Kirui}},\
  }\href@noop {} {\bibfield  {journal} {\bibinfo  {journal} {Physica B:
  Condensed Matter}\ }\textbf {\bibinfo {volume} {613}} (\bibinfo {year}
  {2021})}\BibitemShut {NoStop}%
\bibitem [{\citenamefont {Ongwen}\ \emph {et~al.}(2022)\citenamefont {Ongwen},
  \citenamefont {Ogam}, \citenamefont {Fellah}, \citenamefont {Otunga},
  \citenamefont {Oduor},\ and\ \citenamefont {Mageto}}]{Ongwen2022}%
  \BibitemOpen
  \bibfield  {author} {\bibinfo {author} {\bibfnamefont {N.~O.}\ \bibnamefont
  {Ongwen}}, \bibinfo {author} {\bibfnamefont {E.}~\bibnamefont {Ogam}},
  \bibinfo {author} {\bibfnamefont {Z.}~\bibnamefont {Fellah}}, \bibinfo
  {author} {\bibfnamefont {H.~O.}\ \bibnamefont {Otunga}}, \bibinfo {author}
  {\bibfnamefont {A.~O.}\ \bibnamefont {Oduor}}, \ and\ \bibinfo {author}
  {\bibfnamefont {M.}~\bibnamefont {Mageto}},\ }\href@noop {} {\bibfield
  {journal} {\bibinfo  {journal} {Solid State Communications}\ }\textbf
  {\bibinfo {volume} {353}} (\bibinfo {year} {2022})}\BibitemShut {NoStop}%
\bibitem [{\citenamefont {King'ori}\ \emph {et~al.}(2021)\citenamefont
  {King'ori}, \citenamefont {Ouma}, \citenamefont {Amolo},\ and\ \citenamefont
  {Makau}}]{Kingori2021}%
  \BibitemOpen
  \bibfield  {author} {\bibinfo {author} {\bibfnamefont {G.~W.}\ \bibnamefont
  {King'ori}}, \bibinfo {author} {\bibfnamefont {C.~N.~M.}\ \bibnamefont
  {Ouma}}, \bibinfo {author} {\bibfnamefont {G.~O.}\ \bibnamefont {Amolo}}, \
  and\ \bibinfo {author} {\bibfnamefont {N.~W.}\ \bibnamefont {Makau}},\
  }\href@noop {} {\bibfield  {journal} {\bibinfo  {journal} {Surfaces and
  Interfaces}\ }\textbf {\bibinfo {volume} {24}} (\bibinfo {year}
  {2021})}\BibitemShut {NoStop}%
\bibitem [{\citenamefont {Ouma}\ \emph {et~al.}(2017)\citenamefont {Ouma},
  \citenamefont {Singh}, \citenamefont {Obodo}, \citenamefont {Amolo},\ and\
  \citenamefont {Romero}}]{Ouma201725555}%
  \BibitemOpen
  \bibfield  {author} {\bibinfo {author} {\bibfnamefont {C.~N.}\ \bibnamefont
  {Ouma}}, \bibinfo {author} {\bibfnamefont {S.}~\bibnamefont {Singh}},
  \bibinfo {author} {\bibfnamefont {K.~O.}\ \bibnamefont {Obodo}}, \bibinfo
  {author} {\bibfnamefont {G.~O.}\ \bibnamefont {Amolo}}, \ and\ \bibinfo
  {author} {\bibfnamefont {A.~H.}\ \bibnamefont {Romero}},\ }\href@noop {}
  {\bibfield  {journal} {\bibinfo  {journal} {Physical Chemistry Chemical
  Physics}\ }\textbf {\bibinfo {volume} {19}},\ \bibinfo {pages} {25555 –
  25563} (\bibinfo {year} {2017})}\BibitemShut {NoStop}%
\bibitem [{\citenamefont {Nyawere}\ \emph
  {et~al.}(2014{\natexlab{a}})\citenamefont {Nyawere}, \citenamefont {Makau},\
  and\ \citenamefont {Amolo}}]{Nyawere2014122}%
  \BibitemOpen
  \bibfield  {author} {\bibinfo {author} {\bibfnamefont {P.}~\bibnamefont
  {Nyawere}}, \bibinfo {author} {\bibfnamefont {N.}~\bibnamefont {Makau}}, \
  and\ \bibinfo {author} {\bibfnamefont {G.}~\bibnamefont {Amolo}},\
  }\href@noop {} {\bibfield  {journal} {\bibinfo  {journal} {Physica B:
  Condensed Matter}\ }\textbf {\bibinfo {volume} {434}},\ \bibinfo {pages} {122
  – 128} (\bibinfo {year} {2014}{\natexlab{a}})}\BibitemShut {NoStop}%
\bibitem [{\citenamefont {Ibraheem}\ \emph {et~al.}(2017)\citenamefont
  {Ibraheem}, \citenamefont {Eisa}, \citenamefont {Adlan}, \citenamefont
  {Amolo},\ and\ \citenamefont {Khalafalla}}]{Ibraheem2017}%
  \BibitemOpen
  \bibfield  {author} {\bibinfo {author} {\bibfnamefont {A.}~\bibnamefont
  {Ibraheem}}, \bibinfo {author} {\bibfnamefont {M.}~\bibnamefont {Eisa}},
  \bibinfo {author} {\bibfnamefont {W.}~\bibnamefont {Adlan}}, \bibinfo
  {author} {\bibfnamefont {G.~O.}\ \bibnamefont {Amolo}}, \ and\ \bibinfo
  {author} {\bibfnamefont {M.}~\bibnamefont {Khalafalla}},\ }\href@noop {}
  {\bibfield  {journal} {\bibinfo  {journal} {Modern Physics Letters B}\
  }\textbf {\bibinfo {volume} {31}} (\bibinfo {year} {2017})}\BibitemShut
  {NoStop}%
\bibitem [{\citenamefont {Odhiambo}\ \emph {et~al.}(2015)\citenamefont
  {Odhiambo}, \citenamefont {Amolo}, \citenamefont {Makau}, \citenamefont
  {Dusabirane}, \citenamefont {Othieno},\ and\ \citenamefont
  {Oduor}}]{Odhiambo201569}%
  \BibitemOpen
  \bibfield  {author} {\bibinfo {author} {\bibfnamefont {H.}~\bibnamefont
  {Odhiambo}}, \bibinfo {author} {\bibfnamefont {G.}~\bibnamefont {Amolo}},
  \bibinfo {author} {\bibfnamefont {N.}~\bibnamefont {Makau}}, \bibinfo
  {author} {\bibfnamefont {F.}~\bibnamefont {Dusabirane}}, \bibinfo {author}
  {\bibfnamefont {H.}~\bibnamefont {Othieno}}, \ and\ \bibinfo {author}
  {\bibfnamefont {A.}~\bibnamefont {Oduor}},\ }\href@noop {} {\bibfield
  {journal} {\bibinfo  {journal} {African Review of Physics}\ }\textbf
  {\bibinfo {volume} {10}},\ \bibinfo {pages} {69 – 77} (\bibinfo {year}
  {2015})}\BibitemShut {NoStop}%
\bibitem [{\citenamefont {Odhiambo}\ and\ \citenamefont
  {Othieno}(2015)}]{Odhiambo2015}%
  \BibitemOpen
  \bibfield  {author} {\bibinfo {author} {\bibfnamefont {H.}~\bibnamefont
  {Odhiambo}}\ and\ \bibinfo {author} {\bibfnamefont {H.}~\bibnamefont
  {Othieno}},\ }\href@noop {} {\bibfield  {journal} {\bibinfo  {journal}
  {International Journal of Computational Materials Science and Engineering}\
  }\textbf {\bibinfo {volume} {4}} (\bibinfo {year} {2015})}\BibitemShut
  {NoStop}%
\bibitem [{\citenamefont {Mbilo}\ \emph {et~al.}(2023)\citenamefont {Mbilo},
  \citenamefont {Musembi},\ and\ \citenamefont {Rai}}]{Mbilo20232355}%
  \BibitemOpen
  \bibfield  {author} {\bibinfo {author} {\bibfnamefont {M.}~\bibnamefont
  {Mbilo}}, \bibinfo {author} {\bibfnamefont {R.}~\bibnamefont {Musembi}}, \
  and\ \bibinfo {author} {\bibfnamefont {D.}~\bibnamefont {Rai}},\ }\href@noop
  {} {\bibfield  {journal} {\bibinfo  {journal} {Indian Journal of Physics}\
  }\textbf {\bibinfo {volume} {97}},\ \bibinfo {pages} {2355 – 2362}
  (\bibinfo {year} {2023})}\BibitemShut {NoStop}%
\bibitem [{\citenamefont {Kaner}\ \emph {et~al.}(2023)\citenamefont {Kaner},
  \citenamefont {Wei}, \citenamefont {Fu}, \citenamefont {Ying}, \citenamefont
  {Li}, \citenamefont {Raza}, \citenamefont {Jing}, \citenamefont {Jiang},
  \citenamefont {Meng'wa}, \citenamefont {Yang},\ and\ \citenamefont
  {Li}}]{Kaner20235135}%
  \BibitemOpen
  \bibfield  {author} {\bibinfo {author} {\bibfnamefont {N.~T.}\ \bibnamefont
  {Kaner}}, \bibinfo {author} {\bibfnamefont {Y.}~\bibnamefont {Wei}}, \bibinfo
  {author} {\bibfnamefont {B.}~\bibnamefont {Fu}}, \bibinfo {author}
  {\bibfnamefont {T.}~\bibnamefont {Ying}}, \bibinfo {author} {\bibfnamefont
  {W.}~\bibnamefont {Li}}, \bibinfo {author} {\bibfnamefont {A.}~\bibnamefont
  {Raza}}, \bibinfo {author} {\bibfnamefont {Y.}~\bibnamefont {Jing}}, \bibinfo
  {author} {\bibfnamefont {Y.}~\bibnamefont {Jiang}}, \bibinfo {author}
  {\bibfnamefont {V.~K.}\ \bibnamefont {Meng'wa}}, \bibinfo {author}
  {\bibfnamefont {J.}~\bibnamefont {Yang}}, \ and\ \bibinfo {author}
  {\bibfnamefont {X.}~\bibnamefont {Li}},\ }\href@noop {} {\bibfield  {journal}
  {\bibinfo  {journal} {ACS Applied Energy Materials}\ }\textbf {\bibinfo
  {volume} {6}},\ \bibinfo {pages} {5135 – 5143} (\bibinfo {year}
  {2023})}\BibitemShut {NoStop}%
\bibitem [{\citenamefont {Magnoungou}\ \emph {et~al.}(2022)\citenamefont
  {Magnoungou}, \citenamefont {Malonda-Boungou}, \citenamefont {Amolo},
  \citenamefont {M’Passi-Mabiala},\ and\ \citenamefont
  {Demangeat}}]{Magnoungou2022}%
  \BibitemOpen
  \bibfield  {author} {\bibinfo {author} {\bibfnamefont {J.}~\bibnamefont
  {Magnoungou}}, \bibinfo {author} {\bibfnamefont {B.}~\bibnamefont
  {Malonda-Boungou}}, \bibinfo {author} {\bibfnamefont {G.}~\bibnamefont
  {Amolo}}, \bibinfo {author} {\bibfnamefont {B.}~\bibnamefont
  {M’Passi-Mabiala}}, \ and\ \bibinfo {author} {\bibfnamefont
  {C.}~\bibnamefont {Demangeat}},\ }\href@noop {} {\bibfield  {journal}
  {\bibinfo  {journal} {European Physical Journal B}\ }\textbf {\bibinfo
  {volume} {95}} (\bibinfo {year} {2022})}\BibitemShut {NoStop}%
\bibitem [{\citenamefont {Namisi}\ \emph {et~al.}(2023)\citenamefont {Namisi},
  \citenamefont {Musembi}, \citenamefont {Mulwa},\ and\ \citenamefont
  {Aduda}}]{Namisi2023}%
  \BibitemOpen
  \bibfield  {author} {\bibinfo {author} {\bibfnamefont {M.}~\bibnamefont
  {Namisi}}, \bibinfo {author} {\bibfnamefont {R.~J.}\ \bibnamefont {Musembi}},
  \bibinfo {author} {\bibfnamefont {W.~M.}\ \bibnamefont {Mulwa}}, \ and\
  \bibinfo {author} {\bibfnamefont {B.~O.}\ \bibnamefont {Aduda}},\ }\href@noop
  {} {\bibfield  {journal} {\bibinfo  {journal} {Computational Condensed
  Matter}\ }\textbf {\bibinfo {volume} {34}} (\bibinfo {year}
  {2023})}\BibitemShut {NoStop}%
\bibitem [{\citenamefont {Mbilo}\ and\ \citenamefont
  {Musembi}(2022{\natexlab{b}})}]{Mbilo2022c}%
  \BibitemOpen
  \bibfield  {author} {\bibinfo {author} {\bibfnamefont {M.}~\bibnamefont
  {Mbilo}}\ and\ \bibinfo {author} {\bibfnamefont {R.}~\bibnamefont
  {Musembi}},\ }\href@noop {} {\bibfield  {journal} {\bibinfo  {journal} {AIP
  Advances}\ }\textbf {\bibinfo {volume} {12}} (\bibinfo {year}
  {2022}{\natexlab{b}})}\BibitemShut {NoStop}%
\bibitem [{\citenamefont {Agora}\ \emph {et~al.}(2020)\citenamefont {Agora},
  \citenamefont {Otieno}, \citenamefont {Nyawere},\ and\ \citenamefont
  {Manyali}}]{Agora2020}%
  \BibitemOpen
  \bibfield  {author} {\bibinfo {author} {\bibfnamefont {J.~O.}\ \bibnamefont
  {Agora}}, \bibinfo {author} {\bibfnamefont {C.}~\bibnamefont {Otieno}},
  \bibinfo {author} {\bibfnamefont {P.~W.}\ \bibnamefont {Nyawere}}, \ and\
  \bibinfo {author} {\bibfnamefont {G.~S.}\ \bibnamefont {Manyali}},\
  }\href@noop {} {\bibfield  {journal} {\bibinfo  {journal} {Computational
  Condensed Matter}\ }\textbf {\bibinfo {volume} {23}} (\bibinfo {year}
  {2020})}\BibitemShut {NoStop}%
\bibitem [{\citenamefont {Mbae}\ and\ \citenamefont {Muthui}(2022)}]{Mbae2022}%
  \BibitemOpen
  \bibfield  {author} {\bibinfo {author} {\bibfnamefont {J.~K.}\ \bibnamefont
  {Mbae}}\ and\ \bibinfo {author} {\bibfnamefont {Z.~W.}\ \bibnamefont
  {Muthui}},\ }\href@noop {} {\bibfield  {journal} {\bibinfo  {journal}
  {Advances in Materials Science and Engineering}\ }\textbf {\bibinfo {volume}
  {2022}} (\bibinfo {year} {2022})}\BibitemShut {NoStop}%
\bibitem [{\citenamefont {Nyawere}\ \emph
  {et~al.}(2014{\natexlab{b}})\citenamefont {Nyawere}, \citenamefont
  {Scandolo}, \citenamefont {Makau},\ and\ \citenamefont
  {Amolo}}]{Nyawere201425}%
  \BibitemOpen
  \bibfield  {author} {\bibinfo {author} {\bibfnamefont {P.}~\bibnamefont
  {Nyawere}}, \bibinfo {author} {\bibfnamefont {S.}~\bibnamefont {Scandolo}},
  \bibinfo {author} {\bibfnamefont {N.}~\bibnamefont {Makau}}, \ and\ \bibinfo
  {author} {\bibfnamefont {G.}~\bibnamefont {Amolo}},\ }\href@noop {}
  {\bibfield  {journal} {\bibinfo  {journal} {Solid State Communications}\
  }\textbf {\bibinfo {volume} {179}},\ \bibinfo {pages} {25 – 28} (\bibinfo
  {year} {2014}{\natexlab{b}})}\BibitemShut {NoStop}%
\bibitem [{\citenamefont {Agora}\ \emph {et~al.}(2022)\citenamefont {Agora},
  \citenamefont {Otieno}, \citenamefont {Nyawere},\ and\ \citenamefont
  {Manyali}}]{Agora2022}%
  \BibitemOpen
  \bibfield  {author} {\bibinfo {author} {\bibfnamefont {J.~O.}\ \bibnamefont
  {Agora}}, \bibinfo {author} {\bibfnamefont {C.}~\bibnamefont {Otieno}},
  \bibinfo {author} {\bibfnamefont {P.~W.}\ \bibnamefont {Nyawere}}, \ and\
  \bibinfo {author} {\bibfnamefont {G.~S.}\ \bibnamefont {Manyali}},\
  }\href@noop {} {\bibfield  {journal} {\bibinfo  {journal} {Journal of Physics
  Communications}\ }\textbf {\bibinfo {volume} {6}} (\bibinfo {year}
  {2022})}\BibitemShut {NoStop}%
\bibitem [{\citenamefont {Motochi}\ \emph {et~al.}(2012)\citenamefont
  {Motochi}, \citenamefont {Makau},\ and\ \citenamefont
  {Amolo}}]{Motochi201210}%
  \BibitemOpen
  \bibfield  {author} {\bibinfo {author} {\bibfnamefont {I.}~\bibnamefont
  {Motochi}}, \bibinfo {author} {\bibfnamefont {N.}~\bibnamefont {Makau}}, \
  and\ \bibinfo {author} {\bibfnamefont {G.}~\bibnamefont {Amolo}},\
  }\href@noop {} {\bibfield  {journal} {\bibinfo  {journal} {Diamond and
  Related Materials}\ }\textbf {\bibinfo {volume} {23}},\ \bibinfo {pages} {10
  – 17} (\bibinfo {year} {2012})}\BibitemShut {NoStop}%
\bibitem [{\citenamefont {Philemon}\ and\ \citenamefont
  {Korir}(2020)}]{Philemon2020}%
  \BibitemOpen
  \bibfield  {author} {\bibinfo {author} {\bibfnamefont {K.~T.}\ \bibnamefont
  {Philemon}}\ and\ \bibinfo {author} {\bibfnamefont {K.~K.}\ \bibnamefont
  {Korir}},\ }\href@noop {} {\bibfield  {journal} {\bibinfo  {journal} {Journal
  of Physics Condensed Matter}\ }\textbf {\bibinfo {volume} {32}} (\bibinfo
  {year} {2020})}\BibitemShut {NoStop}%
\bibitem [{\citenamefont {Muthui}\ \emph {et~al.}(2017)\citenamefont {Muthui},
  \citenamefont {Musembi}, \citenamefont {Mwabora},\ and\ \citenamefont
  {Kashyap}}]{Muthui2017343}%
  \BibitemOpen
  \bibfield  {author} {\bibinfo {author} {\bibfnamefont {Z.}~\bibnamefont
  {Muthui}}, \bibinfo {author} {\bibfnamefont {R.}~\bibnamefont {Musembi}},
  \bibinfo {author} {\bibfnamefont {J.}~\bibnamefont {Mwabora}}, \ and\
  \bibinfo {author} {\bibfnamefont {A.}~\bibnamefont {Kashyap}},\ }\href@noop
  {} {\bibfield  {journal} {\bibinfo  {journal} {Journal of Magnetism and
  Magnetic Materials}\ }\textbf {\bibinfo {volume} {442}},\ \bibinfo {pages}
  {343 – 349} (\bibinfo {year} {2017})}\BibitemShut {NoStop}%
\bibitem [{\citenamefont {Chepkoech}\ \emph {et~al.}(2018)\citenamefont
  {Chepkoech}, \citenamefont {Joubert},\ and\ \citenamefont
  {Amolo}}]{Chepkoech2018}%
  \BibitemOpen
  \bibfield  {author} {\bibinfo {author} {\bibfnamefont {M.}~\bibnamefont
  {Chepkoech}}, \bibinfo {author} {\bibfnamefont {D.~P.}\ \bibnamefont
  {Joubert}}, \ and\ \bibinfo {author} {\bibfnamefont {G.~O.}\ \bibnamefont
  {Amolo}},\ }\href@noop {} {\bibfield  {journal} {\bibinfo  {journal}
  {European Physical Journal B}\ }\textbf {\bibinfo {volume} {91}} (\bibinfo
  {year} {2018})}\BibitemShut {NoStop}%
\bibitem [{\citenamefont {Zipporah}\ \emph {et~al.}(2018)\citenamefont
  {Zipporah}, \citenamefont {Robinson}, \citenamefont {Julius},\ and\
  \citenamefont {Arti}}]{Zipporah2018}%
  \BibitemOpen
  \bibfield  {author} {\bibinfo {author} {\bibfnamefont {M.}~\bibnamefont
  {Zipporah}}, \bibinfo {author} {\bibfnamefont {M.}~\bibnamefont {Robinson}},
  \bibinfo {author} {\bibfnamefont {M.}~\bibnamefont {Julius}}, \ and\ \bibinfo
  {author} {\bibfnamefont {K.}~\bibnamefont {Arti}},\ }\href@noop {} {\bibfield
   {journal} {\bibinfo  {journal} {AIP Advances}\ }\textbf {\bibinfo {volume}
  {8}} (\bibinfo {year} {2018})}\BibitemShut {NoStop}%
\bibitem [{\citenamefont {Dongho~Nguimdo}\ \emph {et~al.}(2016)\citenamefont
  {Dongho~Nguimdo}, \citenamefont {Manyali}, \citenamefont {Abdusalam},\ and\
  \citenamefont {Joubert}}]{Nguimdo2016}%
  \BibitemOpen
  \bibfield  {author} {\bibinfo {author} {\bibfnamefont {G.}~\bibnamefont
  {Dongho~Nguimdo}}, \bibinfo {author} {\bibfnamefont {G.~S.}\ \bibnamefont
  {Manyali}}, \bibinfo {author} {\bibfnamefont {M.}~\bibnamefont {Abdusalam}},
  \ and\ \bibinfo {author} {\bibfnamefont {D.~P.}\ \bibnamefont {Joubert}},\
  }\href@noop {} {\bibfield  {journal} {\bibinfo  {journal} {European Physical
  Journal B}\ }\textbf {\bibinfo {volume} {89}} (\bibinfo {year}
  {2016})}\BibitemShut {NoStop}%
\end{thebibliography}%
\end{document}